\documentclass{mn2e}
\usepackage{graphicx}

\title[Blue straggler production in globular clusters]
{Blue straggler production in globular clusters}
\author[Melvyn B. Davies, Giampaolo Piotto, Francesca De Angeli]{
Melvyn B. Davies$^1$, Giampaolo Piotto$^2$, Francesca De Angeli$^2$\\
$^1$ Department of Physics \& Astronomy, University of Leicester,
Leicester, LE1~7RH, UK\\
$^2$ Dipartimento di Astronomia, Universit\`a di Padova, 
Vicolo dell'Osservatorio, 2, I-35122 Padova, Italy}

\begin{document}
\maketitle

\begin{abstract}
Recent HST observations of a large sample of globular
clusters reveal that every cluster contains between 40 and 400 blue stragglers.
The population does not correlate
with either stellar collision rate (as would be expected if all
blue stragglers were formed via collisions) or total mass (as would
be expected if all blue stragglers were formed via the unhindered
evolution of a subset of the stellar population). In this paper,
we support the idea that blue stragglers are made through {\it both} channels.
The number produced via collisions tends to increase with cluster
mass. In this paper we show how the current population produced 
from primordial binaries 
{\it decreases} with increasing cluster mass; exchange encounters with third,
single, stars in the most massive clusters tend to reduce the fraction
of binaries containing a primary close to the current turn-off mass.
Rather their primaries tend to be somewhat more massive ($\sim 1 - 3$ 
M$_\odot$) and have evolved off the main sequence, filling their Roche
lobes in the past, often converting their secondaries into blue stragglers
(but more than 1 Gyr or so ago and thus they are no longer visible as blue 
stragglers). We show that this decline in the primordial blue
straggler population is likely to be offset by the increase in the number
of blue stragglers produced via collisions. The predicted
total blue straggler population
is therefore relatively independent of cluster mass, thus matching
the observed population. This result does not depend on any particular
assumed blue straggler lifetime.
\end{abstract}

\section{Introduction}
Globular Clusters (GCs) are populated by a number of ``exotic'' objects
that cannot be explained by canonical stellar evolution models. Many
of these objects are the results of a number of dynamical interactions
that can modify the original structure of stars, particularly in the
central regions of the GCs, where the dynamical time scale are often
much shorter than the cluster lifetime.

Among these exotic objects there are the blue straggler stars (BSs).
BSs were first identified as an unusual subclass of stars in the
cluster M3 (Sandage 1953). Presently, we know that BSs are present in
all the GCs (Piotto et al. 2003). The leading explanation for their
formation involves mass transfer in and/or the merger of a binary star
system, or collisions between stars (whether or not in binary systems).
A number of photometric studies showed that probably all the suggested
mechanisms of BS formation are at work in different clusters (Fusi
Pecci et al. 1992, Piotto et al. 1999, Ferraro et al. 2003), or
even within the same cluster (Ferraro et al. 1997) .

Very recently, Piotto et al. (2003) have collected a sample of about
3000 BSs, extracted from the colour-magnitude diagrams of 56 GCs, from
their HST snapshot database (Piotto et al. 2002).

The main result of the Piotto et al. (2003) study is a strong
anticorrelation between the BS frequency 
$F_{\rm BS}$\footnote{$F_{\rm BS}=N_{\rm BS}/N_{\rm HB}$, 
where $N_{\rm BS}$ is the number of BSs and $N_{HB}$ is the number of
HB stars identified in the same field}, with the cluster total
luminosity $M_V$ and with the stellar collision rate. Clusters with
higher total luminosity (mass) and clusters with higher collision
probability have a much smaller fraction of BSs.

In this paper, we offer explanations for the 
observations of Piotto et al. (2003).
We show that the observed BS frequency can be the result of 
the combination of the two formation processes, i.e. collisions or mergers 
in a crowded place (which
we call {\it dynamical blue stragglers}) and evolution of
isolated binaries, which we call {\it primordial blue stragglers}.
This result agrees with that of Hurley et al (2001) who concluded
that half of the blue stragglers in M67 have primordial progenitors
whilst half have been formed via dynamical interactions.

In Section 2, we discuss the three observational facts which come
out from the analysis of 
Piotto et al. (2003) and that are of relevance
to this paper.  In Section 3, we consider the relationship between
stellar collision rate and total cluster mass, and compute the likely
size of the dynamical blue straggler population.  We consider the
production of primordial blue stragglers in Section 4, and show why
their production rate {\it decreases} in more--crowded
clusters. We compute the total blue straggler population in Section 5,
and show that the decline in the primordial population is at least
partially offset by the increase in the formation rate of dynamical
blue stragglers in more massive clusters. The
combination of the two processes may well reproduce the population
seen in Piotto et al (2003).

In Section 6 we discuss the effect of mass segregation
on the likely location of the blue straggler population and show that,
at least in some clusters, a large fraction of any primordial 
blue straggler population is still to be found in the cluster halos.

\section{The observational scenario}

Piotto et al. (2003) have shown that there is an anticorrelation
between the BS frequency and the globular cluster
 central collision rate and total
mass (cf. their Fig. 2 and 3).  Indeed, this anticorrelation is the
consequence of three empirical results that we want to explicitly
show here, as they are of relevance to this paper.  As shown in
Figs. 1, and 2 :

\medskip

\noindent A) 
The total number of horizontal--branch (HB) stars in a cluster,
$N_{\rm hb}$ scales as expected with $M_{\rm v}$ such that
$N_{\rm hb} \propto M_{\rm tot}$ (Fig. 1). The total number of
HB and BS stars of Fig. 1 has been estimated by dividing the
number of observed stars by the fraction of cluster total light
covered by the WFPC2 images from which the BSs have 
been extracted.

\medskip

\noindent B) There is a correlation between $M_{\rm tot}$
and the current stellar collision rate, $\Gamma_{\rm coll}$ 
\footnote{$\Gamma_{\rm coll}=5\times10^{-15}(\Sigma_0^3r_c)^{1/2}$, 
where $\Sigma_0$ is the central surface brightness in units of 
$L_{\odot V}$pc$^{-2}$ (equivalent to $\mu_v=26.41$), and $r_c$ 
is the core radius in pc} (Fig. 2).  

\medskip

\noindent C) The total number of blue stragglers in each cluster,
$N_{\rm bs}$ is largely independent of both total mass and 
stellar collision rate (Fig. 1).

\bigskip

The combination of all three of the results listed above produces the
anticorrelation seen in the plot of $F_{BS}$ as a function of
$\Gamma_{\rm coll}$ shown in Fig. 2 of Piotto et al. (2003).  Result
A shows that the number of horizontal branch (HB) stars
scales as the cluster total population, and therefore they must be the result of
the ordinary evolution of stars.
In this paper we consider the relationship between total cluster
mass and current collision rate. We find that the observed
correlation between $M_{\rm v}$
(i.e. total mass) and the current collision rate $\Gamma_{\rm coll}$
(i.e. result B above) is reasonable.
Result C is the hardest to explain. We see from Fig. 1 
that $N_{\rm bs}$ only varies by a factor of about 10 despite the
much larger variations in both $\Gamma_{\rm coll}$ and
$M_{\rm tot}$. If all blue stragglers are formed via collisions then
we would have expected to find $N_{\rm bs} \propto \Gamma_{\rm coll}$. 
There is, in fact, a mild correlation between the total number of blue
stragglers and the collision rate as shown in Fig. 1, but the correlation is
not as strong as would be expected if all blue stragglers are formed
via collisions.
Alternatively, if all blue stragglers are somehow derived from the
initial stellar population (eg. from original binaries), we would have
expected that $N_{\rm bs} \propto M_{\rm tot}$.

   \begin{figure}
   \centering
   \resizebox{\hsize}{!}{\includegraphics{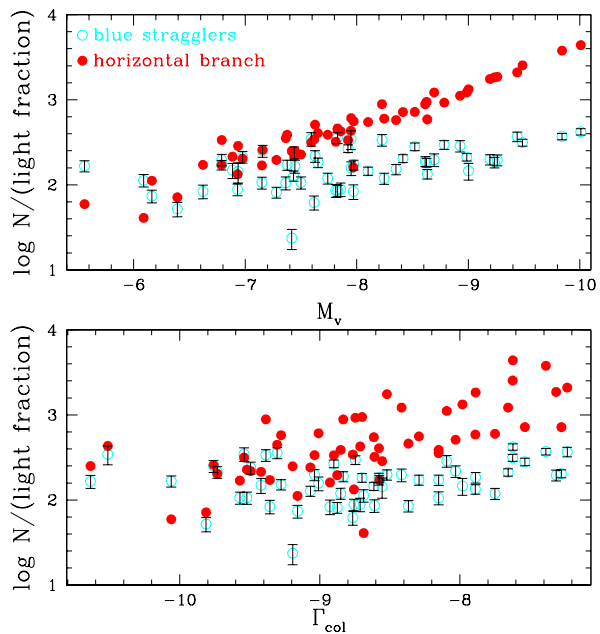}}
   \caption{The estimated total number (see text) of blue straggler  and
   horizontal branch stars in the
   sample of 56 globular clusters as a function of 
   the cluster total magnitude $M_{\rm v}$ ({\it top panel}) and 
   the stellar collision rate, $\Gamma_{\rm col}\equiv5\times10^{-15}
   (\Sigma_0^3r_c)^{1/2}$ in units of collisions yr$^{-1}$
   ({\it bottom panel}).
    See Piotto et al. 2003 for more details.}

\end{figure}

   \begin{figure}
   \centering
   \resizebox{\hsize}{!}{\includegraphics{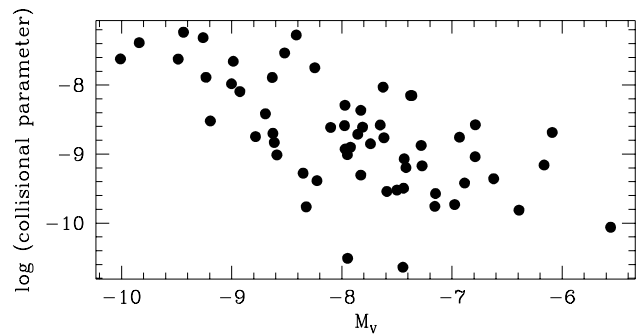}}
   \caption{The cluster total magnitude $M_{\rm V}$ is plotted
against the stellar collision rate $\Gamma_{\rm col}$.}
   \end{figure}


\section{Dynamical blue stragglers}

In this section we compute the size of the blue straggler population
today formed via collisions using the current collision rate,
$\Gamma_{\rm coll}$. 
In the case of a cluster very close to core collapse,
using $\Gamma_{\rm coll}$ may overestimate the number of blue stragglers
which should be observable today.
The collision rate may change
considerably over the lifetime of currently--observed blue stragglers.
This effect is only important in the very few clusters in the
process of core collapse today, and does not affect the blue straggler
population seen in the globular cluster population taken as a whole.
Indeed, Piotto et al. (2003) show that the post--core--collapse
clusters have blue straggler frequencies comparable to the
normal King--model clusters.

The stellar collision rate within the cluster core is given by
$\Gamma_{\rm coll} \propto {\rho^2 r_{\rm c}^3 / \sigma}$, where
$\rho$ is the mass density of stars within the cluster core, $r_{\rm
c}$ is the core radius, and $\sigma$ is the velocity dispersion of the
stars which is $\propto \sqrt{M_{\rm tot}/r_{\rm h}}$. Also $M_{\rm c}
\propto \rho r_{\rm c}^3$. Hence we have

\begin{equation}
\Gamma_{\rm coll} \propto {\rho^2 r_{\rm c}^3 \over \sigma}
\propto {\rho^2 r_{\rm c}^3 \over \sqrt{M_{\rm tot}/r_{\rm h}} }
\propto {M_{\rm c}^2 r_{\rm c}^{-3} \over \sqrt{M_{\rm tot}/r_{\rm h}} }
\propto { f_{\rm c}^2 r_{\rm h}^{1/2} \over r_{\rm c}^3 } M_{\rm tot}^{3/2}
\end{equation}
where $f_{\rm c} = M_{\rm c} / M_{\rm tot}$. Assuming for simplicity
$f_{\rm c}$, $r_{\rm c}$, and $r_{\rm h}$ are the same for all
clusters, we see that $\Gamma_{\rm coll} \propto M_{\rm tot}^{1.5}$.
Clearly $f_{\rm c}$, $r_{\rm c}$, and $r_{\rm h}$ all vary
between clusters, though this simply produces a spread around the
relationship. Hence we can understand why $\Gamma_{\rm coll}$ is seen
to be strongly correlated with $M_{\rm tot}$ in the observed
properties of the globular clusters.

   \begin{figure}
   \centering
   \resizebox{\hsize}{!}{\includegraphics{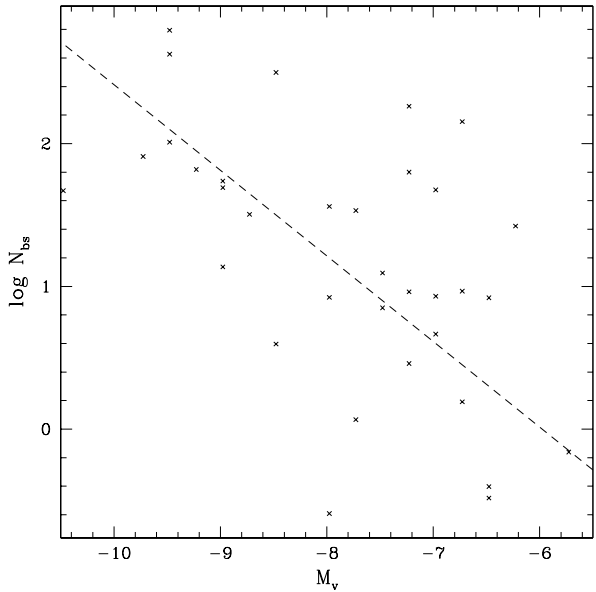}}
   \caption{The number of blue stragglers produced via collisions
during the last 10$^9$ yr as a function of absolute cluster luminosity,
 $M_{\rm v}$, assuming $M/L_{\rm v}=3$ for all clusters and 
$f_{\rm bs} = 0.25$ (see section 3 for details).}
   \end{figure}

We now calculate the number of blue stragglers, formed via collisions,
which are likely to be visible today.
The current collision rate is given by

\begin{equation}
\Gamma_{\rm coll} = N_{\rm c} n_{\rm c} \Sigma_{\rm coll} v_\infty
\end{equation}
where $n_{\rm c}$ is the number density of stars in the core, and
$N_{\rm c}$ is the total number of stars in the core.
The collision cross section, $\Sigma_{\rm coll}$,  is given by

\begin{equation}
\Sigma_{\rm coll} = \pi r_{\rm coll}^2 \left( 1 + { v^2 \over v_\infty^2 }
 \right)
\end{equation}
where $v^2 = 4 G m_\star / r_{\rm coll}$ (where we are assuming all
stars have the same mass, $m_\star$ and $r_{\rm coll}$ is the 
minimum separation of the two stars).
The number of blue stragglers produced via collisions 
during the previous $10^9$ yr is given by:

\begin{equation}
N_{\rm coll,bs}= 0.03225 \left( { f_{\rm mms}^2 N_{\rm c}
n_{\rm c,5} r_{\rm coll}
m_{\rm bs} \over v_\infty} \right)
\end{equation}
where $r_{\rm coll}$ and $m_{\rm bs}$ are in solar units, $n_{\rm c,5}$
is the stellar number density in units of 10$^5$ stars/pc$^3$,
and $f_{\rm mms}$ is the fraction of stars in the core which are massive
main--sequence stars (such that when two collide, they form a blue straggler).
The value of $f_{\rm mms}$ is uncertain, but adopting a value of
$f_{\rm mms} = 0.25$ (see eg Davies \& Benz 1995),
 we plot $N_{\rm coll,bs}$ as a function of total
cluster mass in Figure 3. From this figure we can see that collisions
seem likely to produce $\sim$ 10 -- 100 blue stragglers in a number of
clusters. It should be noted that encounters between binaries and other
binaries/single stars also lead to collisions, but the numbers
will be of the same order. Also included in this figure is a dashed
line following $N_{\rm coll,bs} \propto M_{\rm tot}^{1.5}$; the
relationship derived in equation (1).  $f_{\rm mms}$ is likely to vary
by at least
a factor of 2--3 between clusters, with more concentrated clusters
tending to have larger values of $f_{\rm mms}$ 
as the stars have undergone a greater degree of mass
segregation (see eg Davies \& Benz 1995). This has a significant
effect on $N_{\rm coll,bs}$.  There is also some uncertainty concerning
the lifetime of blue stragglers. In equation (4) we have
in effect assumed a blue straggler lifetime of 1 Gyr. The recent literature
predicts quite a broad range of lifetimes; the spread being somewhat connected
with the treatment of any excess angular momentum obtained via off--axis
collisions (Sills et al 2001; Lombardi et al 2002).
 Even with all these uncertainties, we
are able to conclude that:
{\it in clusters having the highest collision rates,
blue straggler formation via collisions seems likely to produce
a current population of $\sim$ 10--100 blue stragglers}.

\section{Primordial blue stragglers}

In this section we consider the formation of blue
stragglers through the isolated evolution of binaries
as discussed by Preston \& Sneden (2000) and Carney et al (2001). 
This must be the
mechanism to produce the blue stragglers 
observed in the field: we'll  call them 
{\it primordial blue stragglers} in this paper. 
Preston \& Sneden (2000) and Carney et al (2001) consider that such primordial
blue stragglers are formed in relatively {\it wide}
binaries. Rather than having the two stars merge via angular
momentum loss, in these wider systems a blue straggler is
formed when the primary evolves off the main sequence
and fills its Roche lobe. Mass transfer onto the secondary (which
is still a main--sequence star) may then produce a blue straggler.
Certainly very tight binaries will merge, but Preston \& Sneden
argue that this represents a much smaller population.

How would clusterness affect the evolution of primordial blue stragglers?
One might wonder whether the binaries would simply be broken
up by encounters.  This is unlikely, as most are hard (ie binding
energy larger than the kinetic energy of any incoming star).  One
might also wonder whether exchange encounters could alter the
system. For example, if a neutron star exchanged into the system, we
would have a low--mass X--ray binary on mass transfer. This 
certainly occurs, but probably not in sufficient numbers to be the
whole story.  Clusterness produces one very important effect
though, which is shown in Figure 4.  The idea is based on a result
obtained to explain millisecond pulsar production in
globulars (Davies
\& Hansen 1998).
In clusters having high collision rates,
exchange encounters produce binaries containing
more--massive main--sequence primaries.
Once they evolve, mass transfer onto the secondaries produce
blue stragglers, but they will have been formed earlier than in
clusters having low collision rates (where the primary masses are lower).
By today, most of the blue stragglers formed in this way in high collision
rate clusters have evolved, hence we see fewer blue stragglers
in them compared to less--crowded clusters.

   \begin{figure}
   \centering
   \resizebox{\hsize}{!}{\includegraphics{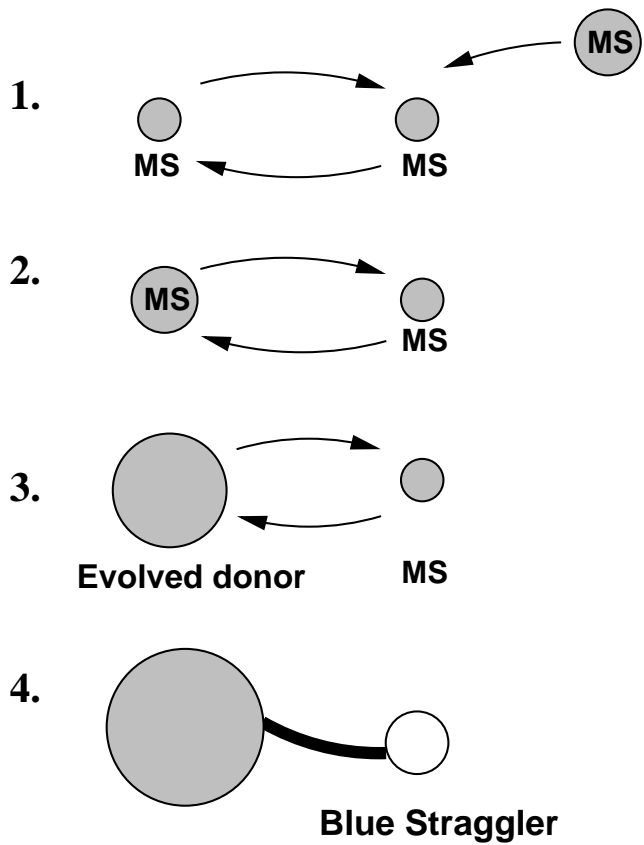}}
   \caption{The evolutionary pathway to produce blue stragglers through
mass--transfer in wide binaries in globular clusters.
A more--massive main--sequence star exchanges into a binary containing
two main--sequence stars (phase 1). The typical primary mass after encounters
in a sufficiently--crowded cluster is $M_1 \simeq 1.5 - 3$ M$_\odot$
(Davies \& Hansen 1998).
This primary evolves of the main--sequence and fills its Roche
Lobe (phase 3). The secondary gains mass from the primary becoming a
blue straggler (phase 4) at a time roughly equal to the main--sequence
lifetime of the donor star. Hence blue stragglers have formed
earlier in binaries containing more--massive primaries (ie in high
collision rate clusters). Given the finite lifetime of blue stragglers,
the blue straggler population in the most crowded clusters today could
be lower than in very sparse clusters.}
              \label{FigGam}%
    \end{figure}

   \begin{figure}
   \centering
   \resizebox{\hsize}{!}{\includegraphics{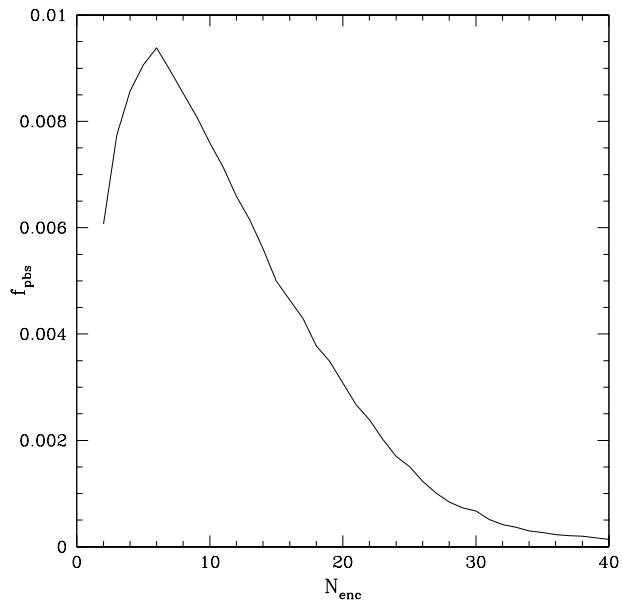}}
   \caption{The fraction of binaries containing a primary of mass between
$0.8$ M$_\odot$ and $0.816$ M$_\odot$
as a function of the number of encounters the binary
undergoes with single stars, $N_{\rm enc}$. Assuming a current turn--off mass
of $0.8$ M$_\odot$, these are the stars which have evolved off the main
sequence within the last 1 Gyr.}
   \end{figure}

To quantify the effect decribed above, we consider the following
simple prescription. We place two stars, drawn from an initial
mass function (IMF, see below), in some initial binary. We then
consider encounters between this binary and a third star, again drawn
from the same IMF. In such encounters involving relatively wide
binaries, by far the most common outcome is the ejection of the least
massive of the three stars, with the remaining two forming a new
binary (see eg Davies \& Benz 1995).  Considering a population of
binaries, we calculate the fraction of systems whose primary masses
are in the range $0.8$ M$_\odot < M_1 < 0.816$ M$_\odot$. This being
the range of masses which have evolved off the main--sequence within
the last 1 Gyr (assuming a current turn--off mass of $0.8$ M$_\odot$ and using
the single--star evolution code of Hurley et al [2000] with a metallicity,
$Z=0.001$).

We assumed a population of stars with an initial mass function given
by Eggleton, Fitchett \& Tout (1989). The mass of one star belonging
to such a population may be generated using the formula below

\begin{equation}
M = {0.19 x \over (1 - x)^{0.75} + 0.032 (1-x)^{0.25} } \ {\rm M}_\odot
\end{equation}
where $x$ is a random number between 0 and 1. 
We are thus able to generate the masses of all stars within a population
by repeated use of the above equation and a series of random numbers.
The fraction of binaries
containing a primary in the required range, $f_{\rm bs}$, is shown in Fig 5
as a function of the number of encounters, $N_{\rm enc}$.
From this figure we clearly see that $f_{\rm bs}$ decreases
significantly when $N_{\rm enc}$ is greater than 10 -- 20. This is
a reasonable number
of encounters for the most crowded globular
clusters. Clearly, there are a number of uncertainties in such a
calculation. The cluster density and stellar mass distribution 
both vary as a function of time.  The number of encounters are also
a function of the distribution of binary separations.

There is also considerable uncertainty concerning
 the lifetime of blue stragglers.
Here we have considered a value of 1 Gyr. Stellar
evolution modelling predicts lifetimes between
100 Myr and 5 Gyr (see for example Sills et al 2001; Lombardi et al 2002). 
However, we have found that the 
exact lifetime considered has no effect on the shape of the curve in Fig
5, or on the results in later sections of this paper.
The general effect described here is robust, namely that more--massive
stars tend to exchange into binaries, and given enough encounters, the
number of systems containing primary stars close to the turn--off mass
is decreased.  

The most--crowded clusters could indeed have contained a much larger blue
straggler population in the past because the primaries in binaries 
get replaced by systematically
more--massive stars. It could also be that the primary masses are sufficiently
large in binaries in the most--crowded clusters, that the evolution once
the donor fills its Roche lobe is quite different: a more extreme mass
ratio may well produce a common envelope phase rather than conservative
mass transfer onto the envelope of the secondary. 
In both cases, the blue straggler population will be small in such
clusters today.

\section{The total blue straggler population}

In this section we calculate the total number of blue
stragglers visible today
within globular clusters combining the contributions from
both dynamical blue stragglers (as described in section 3) and primordial
blue stragglers (as described in section 4).

Following equation (1), we assume that 
the number of {\it dynamical} blue stragglers 
visible today in a particular globular cluster is given by
\begin{equation}
N_{\rm dbs} = k_{\rm dbs} M_{\rm tot}^{3/2}
\end{equation}
where $k_{\rm dbs}$ is a suitably chosen constant. Here we 
assume that $N_{\rm dbs} = 60$ for $M_{\rm tot} = 10^6$ M$_\odot$
(see for comparison, Figure 3, where $M_{\rm tot} = 10^6$ M$_\odot$
is equivalently to $M_{\rm v} = -9.0$).

   \begin{figure}
   \centering
   \resizebox{\hsize}{!}{\includegraphics{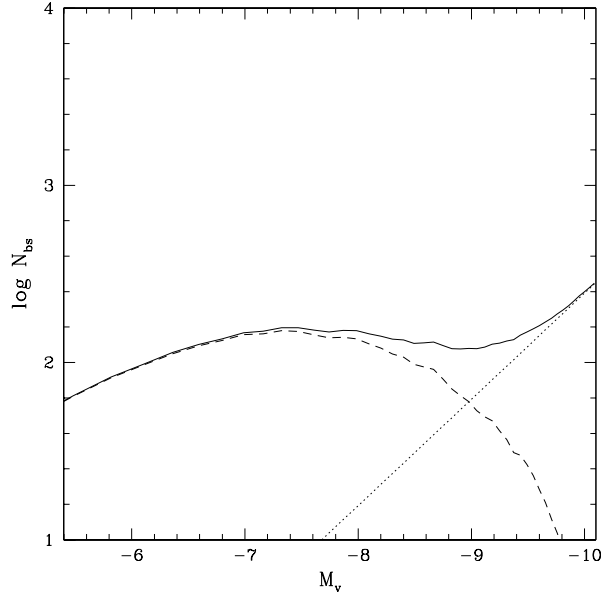}}
   \caption{The number of blue stragglers produced over the last 1 Gyr
as a function of absolute cluster luminosity, $M_{\rm v}$, assuming
$M/L_{\rm v} = 3$ for all clusters. The contribution from primordial systems
is shown with a dashed line, whilst those produced via collisions (involving
either two single stars or binaries) is shown as a dotted line. The total
is given as a solid line.}
   \end{figure}

The number of {\it primordial} blue stragglers 
visible today in a particular globular cluster is given by

\begin{equation}
N_{\rm pbs} = f_{\rm pbs} N_{\rm bin}
\end{equation}
where $f_{\rm pbs}$ is the fraction of binaries containing a primary of
the required mass (as plotted in Figure 5), and $N_{\rm bin}$ is the number
of binaries in the given cluster. 
Here we assume that 
$N_{\rm bin} = 8 \times 10^4$ for $M_{\rm tot} = 10^6$ M$_\odot$.
This may, at first, seem to be rather on the low side, but one
should recall that the primordial blue stragglers are
believed to form from binaries
having a somewhat restricted range of separations (see Preston \& Sneden 2000),
and this reduces the number of potentially relevant binaries by a factor
of a few. The binary fraction tends to be relatively higher for lower--mass
clusters as more binaries are, on average, resilient to breakup. The
separation of binaries lying on the hard--soft boundary, $a_{\rm hs}
\propto \sigma^2$. For clusters having similar radii, the velocity
dispersion within the cluster, $\sigma \propto \sqrt{M_{\rm tot}}$.
If the binaries are distributed uniformly in log(separation) (ie
equal numbers of binaries per decade), then the binary fraction
will scale linearly with log$(M_{\rm tot})$.
We make this assumption here with $N_{\rm bin} = 15 \times 10^3$ for 
$M_{\rm tot} = 10^5$ M$_\odot$.

We also need to relate the number of encounters a binary is likely to have
 to the total cluster mass. As the total number of encounters
within the {\it entire} cluster $\propto M_{\rm tot}^{3/2}$, the number
of encounters one binary is likely to have is $N_{\rm enc}
\propto M_{\rm tot}^{3/2}/N_{\rm bin}  \propto M_{\rm tot}^{1/2}$. 
The timescale for a given binary in a globular cluster
to undergo an encounter with a third star may be approximated as (Binney \&
Tremaine 1987)

\begin{eqnarray}
\tau_{\rm enc} & \simeq & 7 \times 10^{10} {\rm yr} \left( {10^5 \ {\rm pc}^{-3} \over
n } \right) \left( { V_\infty \over 10 \ {\rm km/s} } \right)
\ \nonumber \\ && \times
\left( { {\rm R}_\odot \over R_{\rm min} } \right) \left( { {\rm
M}_\odot \over M } \right) ~,\ 
\end{eqnarray}

\noindent
where $n$ is the number density of single stars and $M$ is
the combined mass of the binary and a typical single star. Taking 
$n = 3 \times 10^4$ stars pc$^{-3}$, $V_\infty = 10$ km/s,
$M = 2.5$ M$_\odot$, $R_{\rm min} = 200$ R$_\odot$, 
we find $\tau_{\rm enc} \simeq 420$ Myr, or in other words
$N_{\rm enc} \simeq 30$, assuming a cluster age of 14 Gyr.
 We assume here that these are values typical
of the most--massive clusters, ie for $M_{\rm tot} = 10^6$ M$_\odot$.

The combined blue straggler populations are shown in Figure 6. To aid
direct comparison to Fig. 1, we have converted total cluster mass to
absolute visual magnitude, assuming $M/L_{\rm v} = 3$ for all
clusters.  The agreement between the population predicted by the
theory and the actual observations shown in Fig. 1 is very
encouraging. 
More important than the absolute numbers is the shape of
the solid curve plotted in Figure 6; the declining primordial
population combines with the increasing dynamical population to
produce a population whose {\it size is relatively independent of
cluster absolute magnitude (and thus mass)}. This is the main result
of this paper. This result is robust to changes in the assumed blue straggler
lifetimes. It also changes very little if we change our assumptions
concerning the binary fraction. If we assume that the binary fraction,
$f_{\rm bin} = 0.08$ for all clusters, then $N_{\rm bs}$ will
decrease by a factor of two (ie $\delta {\rm log}(N_{\rm bs}) 
\simeq 0.3$) for the faintest clusters plotted in Fig 6 whilst
the population in the brightest clusters will be largely unchanged.

Piotto et al (2003) split their observed globular clusters
into two groups: those with $M_{\rm v} > -8.8$ and those with
$M_{\rm v} < -8.8$. From Figure 6, we would predict
that the blue stragglers in the more massive (and thus brighter)
clusters, would be produced predominantly via collisions.
Such blue stragglers are expected to be brighter than those
formed from primordial binaries (Bailyn \& Pinsonneault 1995, but see
also Figure 8 in Piotto et al 1999). 
Indeed Piotto et al (2003, their Fig. 4) find that the luminosity functions
for more--massive clusters do favour brighter blue stragglers, providing
further support to the picture suggested here.

\section{Mass segregation}

Heavy stars sink within the potential well of a cluster on a timescale
given by the local relaxation time. Stars very far out in the halo of a cluster
therefore take much longer to sink into the core. In some clusters
this may prove to be a good observational test of the formation
mechanism of blue stragglers:
in clusters having extremely long
relaxation timescales, any blue stragglers formed from primordial
binaries in the halo of a cluster should still be in the halo, and contained
within wide binaries.

How does the degree of mass segregation scale with $M_{\rm v}$ and
thus $M_{\rm tot}$?  A good indication of what happens
can be obtained by considering the relaxation timescale. The local relaxation
timescale (in years) is given by (see Binney \& Tremaine 1987)

\begin{eqnarray}
t_{\rm r} &=& 0.34 { \sigma^3 \over G^2 m \rho {\rm ln} \Lambda }
\nonumber \\
&=& { 1.8 \times 10^{10} \over {\rm ln} \Lambda }
\left( { \sigma \over 10  {\rm km s}^{-1} } \right)^3
\left( { {\rm M}_\odot} \over m \right)
\left( { 10^3  {\rm M}_\odot}  {\rm pc}^{-3} \over \rho
\right)
\end{eqnarray}
Note that $t_{\rm r} \propto 1/\rho$: the relaxation timescale
is much longer in the lower--density halos of clusters.
We can get some idea of the overall evolution of the cluster
by considering the half--mass relaxation timescale, which is given
by

\begin{eqnarray}
t_{\rm rh} &=& { 0.14 N \over {\rm ln}(0.4 N)} \sqrt{{r_{\rm h}^3
\over G M_{\rm tot}}} \nonumber \\
&=& {6.5 \times 10^8  \ {\rm yr} \over {\rm ln}(0.4 N)}
\left({ M_{\rm tot} \over 10^5 {\rm M}_\odot} \right)^{1/2}
\left( { {\rm M}_\odot} \over m \right)
\left( { r_{\rm h} \over 1 \ {\rm pc} } \right)^{3/2}
\end{eqnarray}
where $m$ is the mass of the particular, heavy star which sinks
in time. Heavier stars tend to take longer to sink in more massive clusters
(the extreme example being Omega Centauri).  Assuming the blue stragglers
are formed from the isolated evolution of primordial binaries, roughly
half of them have sunk to the core in one half--mass relaxation time.
For the globular clusters considered here, 
$t_{\rm rh} \ll 10^{10}$ years in virtually
all cases. However, the relaxation timescale further out than the half--mass
radius is much longer. In some clusters at least, there are
some stars observed where the local relaxation timescale is longer
than the cluster age. Indeed, in
a number of clusters the observations go further out than two half--mass
radii.

If blue stragglers are seen out in the halos, either they are
evolved from isolated primordial binaries, or they have been kicked out
from the core (eg in M3 [Sigurdsson, Davies, Bolte 1994]).
The latter works more often in low--dispersion clusters. 
Information about the radial distribution of the blue stragglers in each
cluster could prove to be extremely useful. 
In clusters having relatively--high velocity dispersions, any blue straggers
found in the halos would have to be derived from primordial binaries.

As an example of the likely population of halo blue stragglers,
we consider an analytic approximation
to a King model as given below (see Hut, McMillan and Romani 1992):

\begin{eqnarray}
\rho &=& \rho_{\rm c} \ \ (r < r_{\rm c}) \nonumber \\
&=& \rho_{\rm c} (r/r_{\rm c})^{-2} \ \ (r_{\rm c}<r<r_{\rm h}) \nonumber \\
&=& \rho_{\rm h} (r/r_{\rm c})^{-4} \ \ (r_{\rm h}<r<r_{\rm t})
\end{eqnarray}
where $\rho_{\rm h} = \rho_{\rm c} (r_{\rm h}/r_{\rm c})^{-2}$
and $\rho_{\rm c} = M_{\rm tot} /[ 8 \pi r_{\rm c}^2 (r_{\rm h} - {2 \over 3}
r_{\rm c})]$. Noting that $t_{\rm rh} \simeq 2 \times 10^9$ yr, for
a number of clusters, we need to consider the mass of the cluster
external to $2 r_{\rm h}$ (where $t_{\rm r} \simeq 10^{10}$ yr,
as $t_{\rm r} \propto 1 / \rho$).
Taking $r_{\rm c}=0.1$ pc, $r_{\rm h} = 5$ pc, and $r_{\rm t} = 50$ pc,
which corresponds to a concentration parameter $c = {\rm log}_{10}
(r_{\rm t}/r_{\rm c})=2.7$, and integrating equation (10)
yields

\begin{equation}
M_{\rm external} = \int_{2r_{\rm h}}^{r_{\rm t}}
4 \pi r^2 \rho dr \simeq M_{\rm tot} / 4
\end{equation}

In other words:
with the above cluster model, 1/4 of any blue stragglers
formed from the isolated evolution of binaries should still be located in
the outer halo.

\section{Conclusions}

The observations of the blue straggler population within
globular clusters by Piotto et al (2003) reveal that the number
of blue stragglers within each cluster is restricted
to a relatively narrow range of values (between 40 -- 400 blue
stragglers per cluster). If the observed blue
straggler population is derived entirely from the unhindered evolution
of primordial binaries (as presumedly must occur in the field), then
we would have expected the number of blue stragglers to scale linearly
with total cluster mass (and thus the blue stragglers population
per cluster would vary by a factor of about 100). Alternatively,
if all blue stragglers are formed via stellar collisions, then we would
expect to see a strong correlation between the size of the blue straggler
population and the collision rate within a given cluster. This is not
observed.

In relatively wide binaries, a blue straggler may be produced 
when the primary evolves off the main sequence, transferring material
onto the secondary which is still on the main sequence. Observations
suggest that this is what produces the blue stragglers seen in the halo
(Carney et al 2001; Sneden et al 2003).
We have dubbed them primordial blue stragglers in this paper. Such binaries
are  vulnerable to exchange encounters in the crowded environments of stellar
clusters. In exchange encounters,  low--mass components of the binary are
replaced by more--massive single stars. The effect is to increase the
average mass of the primary with increasing numbers of encounters. 
The blue straggler population observed today is derived from systems
where mass transfer has occurred in the last 1 Gyr or so (ie where the 
primary has evolved off the main sequence within the last 1 Gyr).  
Encounters between binaries and single stars tend to {\it reduce} the number
of binaries containing primaries with masses close to the present
turn--off mass when the number of encounters exceeds 10 -- 20. Thus
the population of primordial blue stragglers is reduced in more
crowded (and on average, more massive) clusters.

The stellar collision rate is correlated with cluster mass. Hence
the number of dynamically--formed blue stragglers increases with
cluster mass. This increase offsets 
the decline in the population of primordial blue stragglers. The 
predicted total blue  straggler population is relatively uniform for
all cluster masses, reproducing the observed blue straggler populations.

We have also considered the effects of  mass segregation
on the current blue straggler population. In some clusters,
any surviving primordial blue stragglers may still be found in the outer parts
of the halo. 

\bigskip

\section{Acknowledgements}

      MBD acknowledges funding from the Particle Physics and Astronomy
Research Council. GP and FDA acknowledge the support of the Ministero
dell'Universit\`a e della Ricerca (PRIN2001) and of the Agenzia Spaziale
Italiana.

\end{document}